\documentclass[pra,superscriptaddress,twocolumn,superscriptaddress,amsmath]{revtex4}

\usepackage{amsmath, amsthm, amssymb}
\usepackage[]{graphicx} 
\usepackage{dcolumn} 
\usepackage{bm} 
\usepackage[normalem]{ulem}
\usepackage{cases}
\usepackage{color}
\usepackage[dvipsnames]{xcolor}
\usepackage{algorithm}
\usepackage{algpseudocode}
\usepackage{csquotes}
\usepackage{mathtools}
\usepackage{mathrsfs}  
\usepackage{mathtools}
\usepackage[caption=false]{subfig}
\usepackage{lipsum}
\usepackage{comment}
\usepackage{tabularx}
\usepackage{multirow}
\usepackage{braket}
\usepackage{pgf} 
\usepackage{siunitx} 

\usepackage{hyperref}
\hypersetup{
    colorlinks=true,
    linkcolor=blue,
    citecolor=blue,
    filecolor=magenta,      
    urlcolor=cyan
    }

\def\Rx{\textup{Rx}}
\def\sat{\textup{sat}}

\def\MA{\textup{MA}}
\def\OP{\textup{OP}}
\def\PTF{\textup{PTF}}
\def\wtau{\widetilde{\tau}}

\usepackage[colorinlistoftodos]{todonotes}

\begin{document}

\title{International time transfer between precise timing facilities \\ secured with a quantum key distribution network}

\author{Francesco~Picciariello}
\affiliation{Dipartimento di Ingegneria dell'Informazione, Universit\`a degli Studi di Padova, via Gradenigo 6B, IT-35131 Padova, Italy}

\author{Francesco~Vedovato}
\email[Corresponding author: ]{francesco.vedovato@unipd.it}
\affiliation{Dipartimento di Ingegneria dell'Informazione, Universit\`a degli Studi di Padova, via Gradenigo 6B, IT-35131 Padova, Italy}
\affiliation{Padua Quantum Technologies Research Center, Universit\`a degli Studi di Padova, via Gradenigo 6A, IT-35131 Padova, Italy}

\author{Davide~Orsucci}
\author{Pablo~Nahuel~Dominguez}
\author{Thomas~Zechel}
\affiliation{German Aerospace Center (DLR), Institute for Communications and Navigation, M\"unchener Stra{\ss}e 20, 82234 We{\ss}ling, Germany}

\author{Marco~Avesani}
\affiliation{Dipartimento di Ingegneria dell'Informazione, Universit\`a degli Studi di Padova, via Gradenigo 6B, IT-35131 Padova, Italy}

\author{Matteo~Padovan}
\affiliation{Dipartimento di Ingegneria dell'Informazione, Universit\`a degli Studi di Padova, via Gradenigo 6B, IT-35131 Padova, Italy}
\affiliation{Centro di Ateneo di Studi e Attivit\`a Spaziali ``Giuseppe Colombo'', Universit\`a di Padova, via Venezia 15, IT-35131 Padova, Italy}

\author{Giulio~Foletto}
\affiliation{Dipartimento di Ingegneria dell'Informazione, Universit\`a degli Studi di Padova, via Gradenigo 6B, IT-35131 Padova, Italy}

\author{Luca~Calderaro}
\affiliation{ThinkQuantum s.r.l., via della Tecnica 85, IT-36030 Sarcedo, Italy}

\author{Daniele~Dequal}
\affiliation{Telecommunication and Navigation Division, Agenzia Spaziale Italiana, Matera, Italy}

\author{Amita~Shrestha}
\author{Ludwig~Bl\"umel}
\author{Johann~Furthner}
\affiliation{German Aerospace Center (DLR), Institute for Communications and Navigation, M\"unchener Stra{\ss}e 20, 82234 We{\ss}ling, Germany}

\author{Giuseppe~Vallone}
\author{Paolo~Villoresi}
\affiliation{Dipartimento di Ingegneria dell'Informazione, Universit\`a degli Studi di Padova, via Gradenigo 6B, IT-35131 Padova, Italy}
\affiliation{Padua Quantum Technologies Research Center, Universit\`a degli Studi di Padova, via Gradenigo 6A, IT-35131 Padova, Italy}

\author{Tobias~D.~Schmidt}
\author{Florian~Moll}

\affiliation{German Aerospace Center (DLR), Institute for Communications and Navigation, M\"unchener Stra{\ss}e 20, 82234 We{\ss}ling, Germany}

\begin{abstract}

Global Navigation Satellite Systems (GNSSs), such as GPS and Galileo, provide precise time and space coordinates globally and constitute part of the critical infrastructure of modern society. To reliably operate GNSS, a highly accurate and stable system time is required, such as the one provided by several independent clocks hosted in Precise Timing Facilities (PTFs) around the world.
Periodically, the relative clock offset between PTFs is measured to have a fallback system to synchronize the GNSS satellite clocks.
The security and integrity of the communication between PTFs is of paramount importance: if compromised, it could lead to disruptions to the GNSS service.
Therefore, it is a compelling use-case for protection via Quantum Key Distribution (QKD), since this technology provides information-theoretic security.
We have performed a field trial demonstration of such use-case by sharing encrypted time synchronization information between two PTFs, one located in Oberpfaffenhofen (Germany) and one in Matera (Italy) -- more than 900km apart as the crow flies. 
To bridge this large distance, a satellite-QKD system is required, plus a “last-mile” terrestrial link to connect the optical ground station (OGS) to the actual location of the PTF. 
In our demonstration we have deployed two full QKD systems to protect the last-mile connection at both the locations and have shown via simulation that upcoming QKD satellites will be able to distribute keys between Oberpfaffenhofen and Matera exploiting already existing OGSs.

\end{abstract}

\maketitle


\section{Introduction}

\begin{figure*}[t]
    \centering
    \includegraphics[width=\textwidth ]{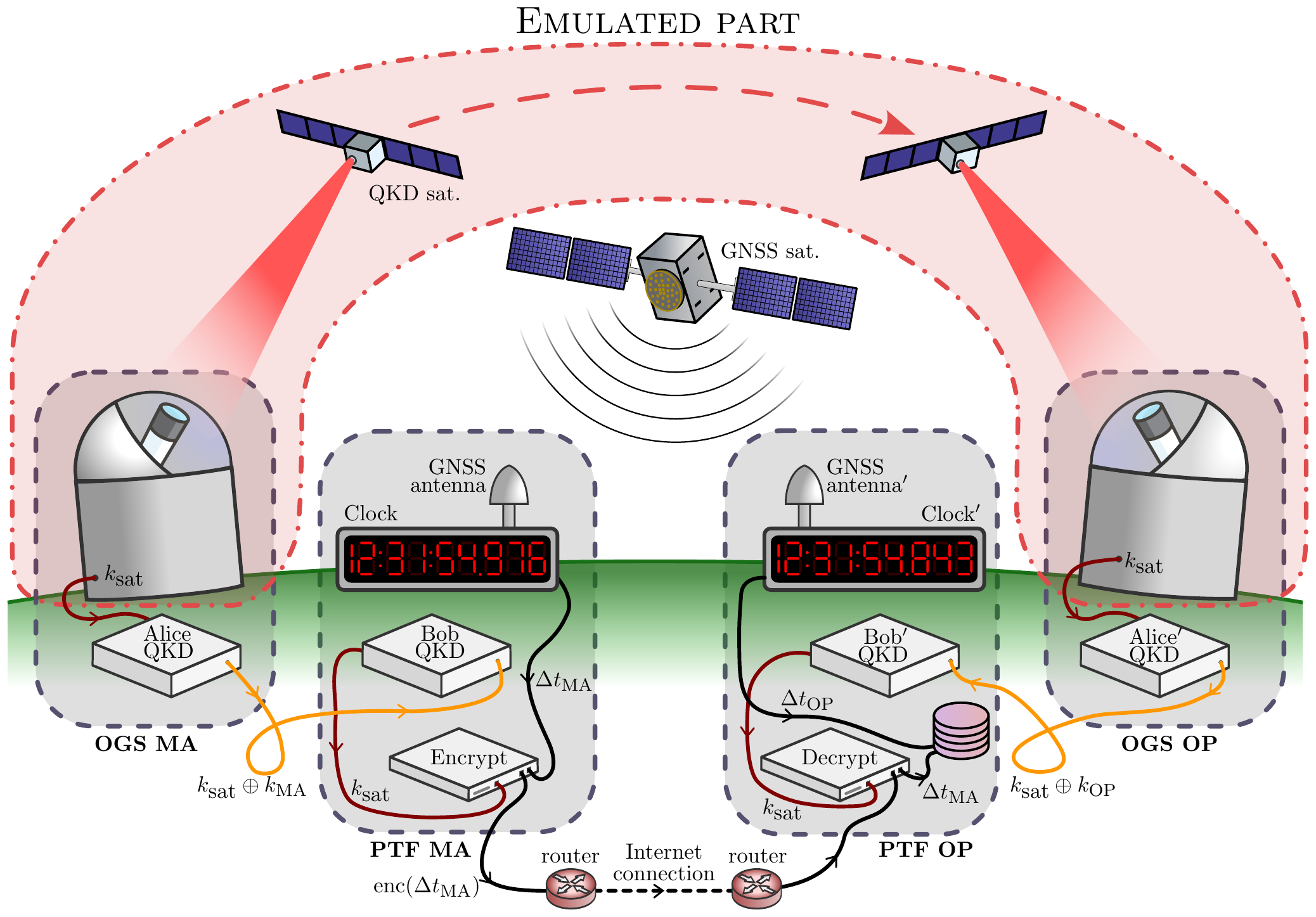}
    \caption{
    Representation of the quantum encrypted time-transfer use-case demonstration. Matera (MA) labs are on the left and Oberpfaffenhofen (OP) labs are on the right. The gray dashed lines enclose the equipment considered to be within secure perimeters (physically inaccessible to attackers). The red dashed-dotted line encloses the equipment that has been emulated and not yet experimentally demonstrated due to the unavailability of QKD satellites. 
    Description: a QKD satellite distributes the key $k_\sat$ to both the OGS in MA and OP; the two last-mile QKD connections generate quantum keys $k_\MA$ and $k_\OP$ which are then used to one-time pad the key $k_\textup{sat}$ and securely forward it to the PTF in MA and OP (transmitting the encryptions $k_\textup{sat} \oplus k_\MA$ and $k_\textup{sat} \oplus k_\OP$, respectively); the time differences $\Delta t_\OP$ and $\Delta t_\MA$ between the local clocks and a GNSS satellite are measured in a all-in-view experiment; the $\Delta t_\OP$ value is AES-encrypted with key $k_\textup{sat}$ and is transmitted to MA over the internet, where it is AES-decrypted using the same key $k_\textup{sat}$; the values $\Delta t_\MA$ and $\Delta t_\OP$ are saved and stored locally in OP, where it can be used to monitor the clock difference $\Delta t_\textup{OP,MA} = \Delta t_\OP - \Delta t_\MA$. } 
    \label{fig:summary}
\end{figure*}

The currently deployed public-key cryptography infrastructure hinges on computational assumptions. These security assumptions are being challenged by the advent of quantum computers, which, by implementing Shor's algorithm~\cite{Shor1994}, can break the most common public-key encryption schemes, such as RSA~\cite{RSA1978} and elliptic curve cryptosystems~\cite{Miller1986}. It is therefore of paramount importance to start transitioning away from the currently employed public-key cryptosystems. Although there exist new cryptographic algorithms that are believed to be resilient to quantum computation attacks~\cite{Bernstein2017}, a more radical and long-term solution is to rely on Quantum Key Distribution (QKD). This technology allows secure communication without using computational assumptions, but rather harnesses the laws of quantum mechanics to achieve information-theoretical security~\cite{BB84,Xu2020,Pirandola:20}. QKD can fully evade the threat posed by the upcoming quantum computers and by all other potential algorithmic and hardware advancements.

The critical services that underpin the digital and communication infrastructures, because of their crucial importance for modern societies, shall be among the first systems to be upgraded to post-quantum-computation security standards through the use of QKD. The European Union is pushing in this direction by developing and testing experimental quantum communication networks in several European countries, through initiatives such as OpenQKD~\cite{OPENQKD}.

In this work, realized within the OpenQKD project, we demonstrate the possibility of securely transmitting clock difference data needed for synchronization of clocks of two distant Precise Timing Facilities (PTFs). One is located at the German Aerospace Center (Deutsches Zentrum für Luft- und Raumfahrt, DLR) in Oberpfaffenhofen (OP), Germany; the second is located at the Matera Laser Ranging Observatory (MLRO) of the Italian Space Agency in Matera (MA), Italy. This is a highly impactful use-case, since attacks performed on the clock synchronization data between PTFs could lead to large-scale service disruptions. In fact, an attacker who can manipulate or forge synchronization data has the possibility to introduce small temporal shifts in the clocks that employ such data for synchronization; these shifts may result in a slow drift of the global system time, which is generated by one of the PTF in duty. Global Navigation Satellite Systems (GNSSs), for instance, require a very precise system time to operate correctly, and an attack to the underlying PTF could result in a worldwide disruption of the service.

In our QKD demonstration, we have secured the transmission of synchronization data between two PTFs that are located more than 900~km apart (but which are not actually part of the Galileo ground segment). This distance is too large to perform a QKD exchange via a direct optical fiber connection, due to the exponential loss of optical power~\cite{Pirandola:20}. A method that could allow bridging such a long distance with currently existing and demonstrated technology would be to employ a satellite to create a QKD link between the two facilities. This may be done either by using a satellite as a trusted node to relay a quantum-generated key~\cite{liao2018} or, more ambitiously, by employing a satellite that can generate and distribute entangled photon pairs simultaneously~\cite{yin2020} to the two PTFs. Both the PTF in MA and the one in OP have an Optical Ground Station (OGS) located nearby that is capable of receiving optical quantum states from a satellite in Low Earth Orbit (LEO) and will be employed, in the future, for satellite-to-ground QKD implementations. 
The European Union is developing technologies for satellite-based QKD, for example, through the construction of the Eagle-1 demonstrator satellite \cite{sidhu2021} and the activities of the Secure And cryptoGrAphic Mission (SAGA) \cite{lewis2022}; a comprehensive review on QKD satellite projects can be found in Ref.~\cite{sidhu2021}. The last step still required is the secure forwarding of the quantum-generated key from the OGS to the PTF at each location. This “last-mile" connection is short enough that it can be secured with a QKD link over an optical fiber, as we experimentally demonstrated in our implementation. 
The general overview of how to secure communication between the two PTFs with a QKD satellite and two last-mile connections is presented schematically in Figure~\ref{fig:summary}. 

The outline of the paper is as follows. In Section~\ref{sec:setup} we give a preliminary description of the use-case demonstration and the experimental setup. In Section~\ref{sec:last-mile} we present the realization of the two last-mile QKD connections and the observed key generation rates. In Section~\ref{sec:satQKD} we present the simulations of a satellite quantum communication link and show that the connection is feasible with existing OGS infrastructure, such as the two OGSs in MA and OP, 
and upcoming QKD satellites. In Section~\ref{sec:GNSS} we give more details on clock data acquisition and on the all-in-view measurement of time offsets. In Section~\ref{sec:conclusion} we present our conclusions and outlooks.


\section{Use-case demonstration description and experiment setup}
\label{sec:setup}

In this section, we give a high-level description of the use-case demonstration, as schematically presented in Figure~\ref{fig:summary}. The demonstration involved the coordinated and simultaneous operation of four experimental parts: (1) a time offset measurement between the PTF clocks in OP and in MA, made possible by all-in-view detection of GNSS signals from the Galileo constellation (2) a fiber-based QKD link in MA (3) a second QKD link in OP and (4) the real-time, encrypted, and authenticated transmission of time difference data from MA to OP.

Notice that, since no European satellite capable of quantum state downlink is currently available, part of the demonstration that would require the use of a satellite has not been performed experimentally. In Section~\ref{sec:satQKD} we show, however, that it is feasible to establish a QKD link with the currently existing OGS, making reasonable assumptions on the performance of the satellite quantum communication system. 

\subsection{Time offset measurement}
\label{sec:common-view}

The time difference between the clock in OP and the clock in MA is measured according to the common-view principle, performed by measuring the time difference between the local clocks and the clock of a GNSS satellite which is in line of sight from both PTFs~\cite{allan1980,DeFraigne2003}.

GNSS satellites transmit their internal clock $t_\sat$ through radio signals. At the PTF in OP (respectively, MA) the local clock $t_\OP$ (respectively, $t_\MA$) is compared with $t_\sat$, acquired with geodetic GNSS receivers. Due to the finiteness of the speed of light, the signal $t_\sat$ arrives at the PTF only after a certain propagation time $\tau_{\sat,\PTF}$ and thus the actual measured time difference is
\begin{equation}
       \Delta t_{\PTF,\sat} = t_\PTF - \big( t_\sat - \tau_{\sat,\PTF} \big)
       \label{eq:deltaT_loc}
\end{equation}
where $t_\PTF$ can be referred to $t_\MA$ or $t_\OP$, and similarly for $\tau_{\sat,\PTF}$ and $\Delta t_{\PTF,\sat}$. The propagation time is estimated using satellite ephemeris data (which are calibrated through satellite ranging measurements~\cite{allan1980,DeFraigne2003}) and the geographic position of the GNSS receiving antenna. Further calibration data, such as cable delays and ionospheric effects, are required to obtain an accurate estimate $\widetilde{\tau}_{\sat,\PTF}$ of $\tau_{\sat,\PTF}$. The corrected time difference is then computed as
\begin{equation}
    \Delta t_{\PTF,\sat}^\textup{corr} = t_\PTF - \big( t_\sat - \tau_{\sat,\PTF} + \wtau_{\sat,\PTF}\big)
\end{equation}
and the time offset between the two PTF clocks is measured by subtracting the two (corrected) time differences
\begin{equation}
    \Delta t_{\OP,\MA}^\textup{meas}
    = \Delta t_{\OP,\sat}^\textup{corr} - \Delta t_{\MA,\sat}^\textup{corr}
    \simeq t_\OP - t_\MA
    \label{eq:deltaT}
\end{equation}
Notice that after subtraction, the dependence on the satellite clock is factored out. 

In an all-in-view measurement, several satellites of one GNSS constellation or of multiple constellations can be employed simultaneously to obtain a more precise estimate of the time offset, with the accuracy increasing roughly as the square root of the number of employed satellites. Furthermore, employing a GNSS constellation allows the measurement of the time offset between two PTFs even if there is never a single satellite in common-view (e.g., if the PTFs are located on antipodal points on Earth). This is possible because the clocks in a GNSS constellation are mutually synchronized. In this scenario, the (corrected) time difference between the PTF to GNSS system time is obtained as
\begin{equation}
    \Delta t_{\PTF,\textup{GNSS}}^\textup{corr} = t_\PTF - \bar t_\textup{GNSS}
    \label{eq:deltaT_loc2}
\end{equation}
where the median
\begin{equation}
    \bar t_\textup{GNSS} = \underset{\sat \in \textup{GNSS}}{\textup{median}} \big(t_\sat - \tau_{\sat,\PTF} + \widetilde{\tau}_{\sat,\PTF} \big)
    \label{eq:median}
\end{equation}
is computed over the set of GNSS satellites that are visible from the PTF in a given moment; the median is employed since it is a more robust estimator than the average (e.g., it is insensitive to glitches in time read-out from a single satellite). Note that ephemeris data for each visible satellite are required. The time offset between OP and MA is obtained as the difference
\begin{equation}
    \Delta t_{\OP,\MA}^\textup{meas} = \Delta t_{\OP,\textup{GNSS}}^\textup{corr} - \Delta t_{\MA,\textup{GNSS}}^\textup{corr} \simeq t_\OP - t_\MA    
    \label{eq:deltaT2}
\end{equation}
whereby the GNSS system time is factored out.

In our demonstration,  clock data from 16 Galileo satellites has been collected over the course of two days. Details on the acquisition of GNSS data and the estimation of the propagation time $\tau_{\sat,\PTF}$ are presented in Section~\ref{sec:GNSS}.

\subsection{Securing the last-mile connections with QKD}
\label{sec:securing}

QKD allows generating a common secret bit-string between distant parties, usually denoted as Alice and Bob, who are linked via a quantum communication channel and a classical authenticated channel. The security is guaranteed by the laws of quantum mechanics: it is impossible to faithfully copy an unknown quantum state, and, thus, an eavesdropper, usually called Eve, that attempts to access the information encoded in a quantum state inevitably introduces excess noise. By monitoring the fidelity of the exchanged quantum states, Alice and Bob can establish rigorous upper bounds to the information available to Eve. If the noise is too high, Eve may have access to too much mutual information, and the protocol has to abort; otherwise, a secure quantum key can be generated. Information reconciliation and privacy amplification, classical post-processing algorithms, are employed to achieve almost perfect privacy and correctness from a raw key~\cite{Xu2020, Pirandola:20}. 

In our use-case, two QKD links have been established using commercial QKD systems provided by ThinkQuantum s.r.l. \cite{TQ} in OP and research prototypes developed by the University of Padova in MA. The purpose of these links is to allow the secure forwarding of information in a last-mile connection in both OP and MA, with Alice modules (QKD transmitters) located at the OGSs and Bob modules (QKD receivers) located at the PTFs. These two last-mile connections are used in a trusted-node configuration: the OGS acts as an intermediate trusted-node that generates  $k_\sat$ and securely forwards it to its final destination at the PTF. The fiber-based QKD link in MA (respectively, OP) is used to distribute a locally-generated key $k_\MA$ ($k_\OP$) between Alice and Bob's modules. This key is then used to one-time-pad the satellite key and the resulting cyphertext $k_\sat \oplus k_\MA$ ($k_\sat \oplus k_\OP$) is forwarded from the Alice module to the Bob module over the authenticated classical channel (the same that is used to support the QKD post-processing). The Bob module employs $k_\MA$ ($k_\OP$) to recover the value $k_\sat$ and finally the key $k_\MA$ ($k_\OP$), having completed its function, is deleted.

As analyzed in Section~\ref{sec:satQKD}, the distribution of a key $k_\sat$ to the OGS in MA and OP is feasible with satellite QKD technology that will be available in the near future. For the purpose of the current demonstration, a random bit string $k_\sat$ had been uploaded into both Alice modules just prior to their shipment to MA and OP, respectively. We note that when a real satellite QKD system is employed, the end-to-end QKD link is established in a similarly asynchronous manner. In fact, the latency in the distribution of quantum keys via satellite can be of several days or weeks, depending on the number of QKD satellites available and on the local weather conditions. Therefore, the key $k_\sat$ must be created well in advance of its intended use and buffered locally for the necessary time.

\subsection{Data encryption, authentication and real-time data transfer}

The final part of the use-case demonstration is to secure the communication of the time offset measurements between MA and OP. We note that these data do not have particular privacy or confidentiality requirements, and thus are not required to be encrypted. However, the communication has to be authenticated in order to avoid data manipulation by potential attackers. 

To secure the communication of the time offset measurements, we exploited a commercial encryptor prototype made by Rhode \& Schwarz GmbH \& Co~\cite{res}, the SITLine ETH4G/40G. Both the SITLines and the four QKD systems provide a QKD-adapted interface, the ETSI GS QKD 004, which enables the communication between the two types of devices to use the quantum secret key. This type of interfaces are fairly new in the QKD field and show the systems' readiness for more advanced QKD networks.

As sketched in Figure~\ref{fig:summary}, every two minutes, the two encryptors ask the Bob QKD devices for a secret key. The refresh time of the key is a security parameter that can be relaxed depending on the needs. The SITLine works at the data-link layer (L2 of the ISO/OSI model), and in order to send the encrypted packets from MA to OP through the Internet we had to implement an encapsulation of the data into the network layer (L3 of the ISO/OSI model) with the two routers depicted in the scheme. The key is 256~bits long, thus allowing the encryptors to implement the AES-256 encryption protocol on all antenna traffic~\cite{AES}. Since the encryptors implement AES by default and we did not have particular restrictions on the key production, we chose to encrypt all the data transferred during the experiment. Note that by using a Wegman-Carter scheme, it could be possible to authenticate the data with information-theoretic security using a number of secret bits that scale only logarithmically in the message length~\cite{WC1981}. 

Once the classical channel between the two routers is enabled, the data acquired from the antenna in MA is automatically sent to the encryptor every 30 minutes, decrypted in OP and stored in a computer for the analysis. Note that our encryption scheme ensures the security of the data from the GNSS antennas to the other end of the routers. However, spoofing of the GNSS signals is a current issue for many satellite positioning systems~\cite{Ceccato:18}, and these types of attack could poison the data at the source. As of now, this is avoided by using the Galileo satellite network, which uses Open Service Navigation Message Authentication (OSNMA), which provides stronger authentication methods than other systems, such as GPS~\cite{OSNMA, OSNMA2021}; furthermore, integrity and consistency checks are performed in the GNSS data, alongside the use of multi-array antennas to reduce spoofing and jamming. In the future, QKD satellites could be integrated into the network to also authenticate this part of the signal~\cite{Dai2020}.

\begin{figure*}[t!]
    \centering
    \begin{minipage}{.49\textwidth}
        \textbf{MA Last-mile QKD link} 
    \end{minipage}
    \begin{minipage}{.49\textwidth}
        \textbf{OP Last-mile QKD link} 
    \end{minipage}
    \includegraphics[width=.49\textwidth ]{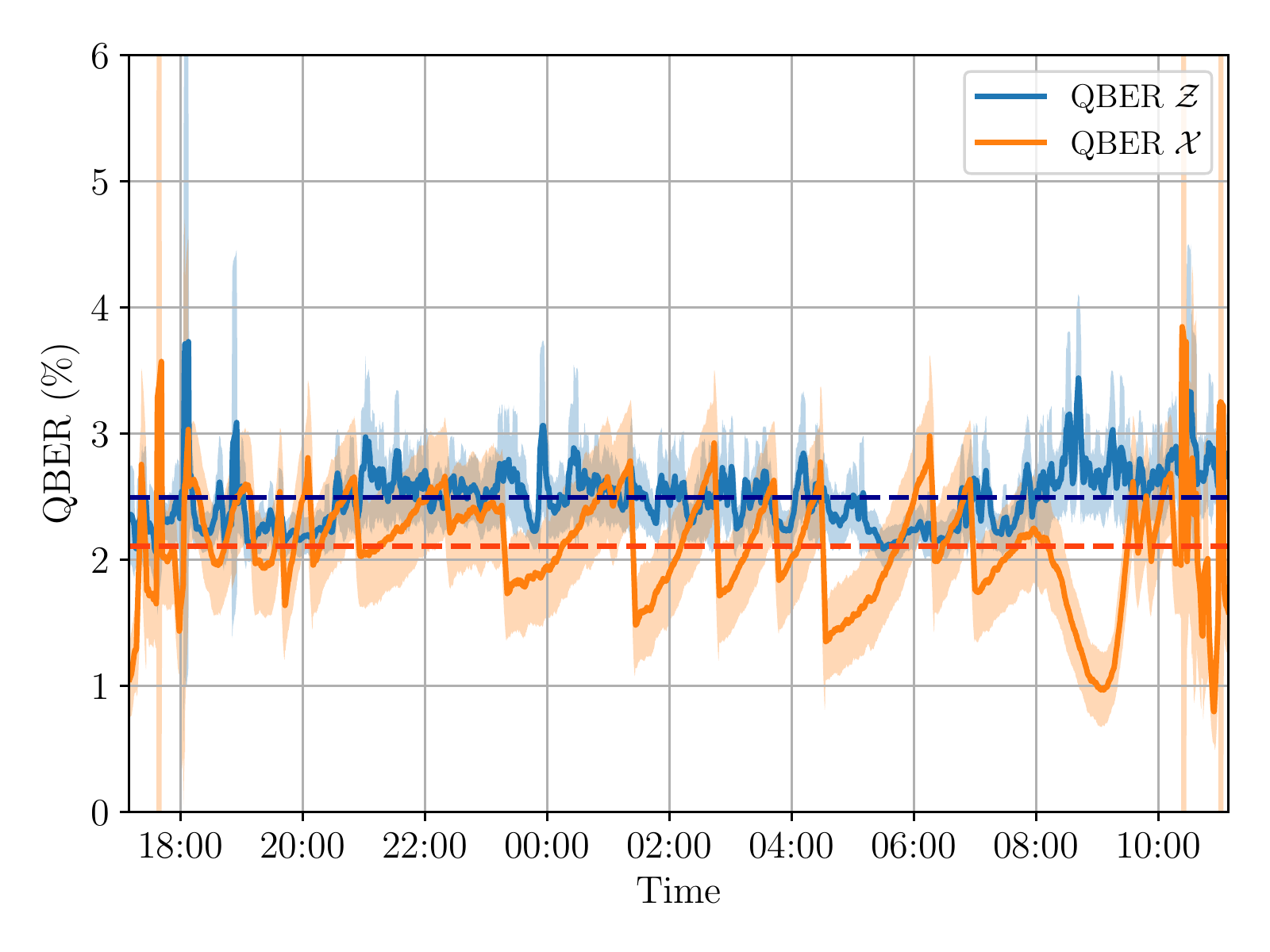}
    \includegraphics[width=.49\textwidth ]{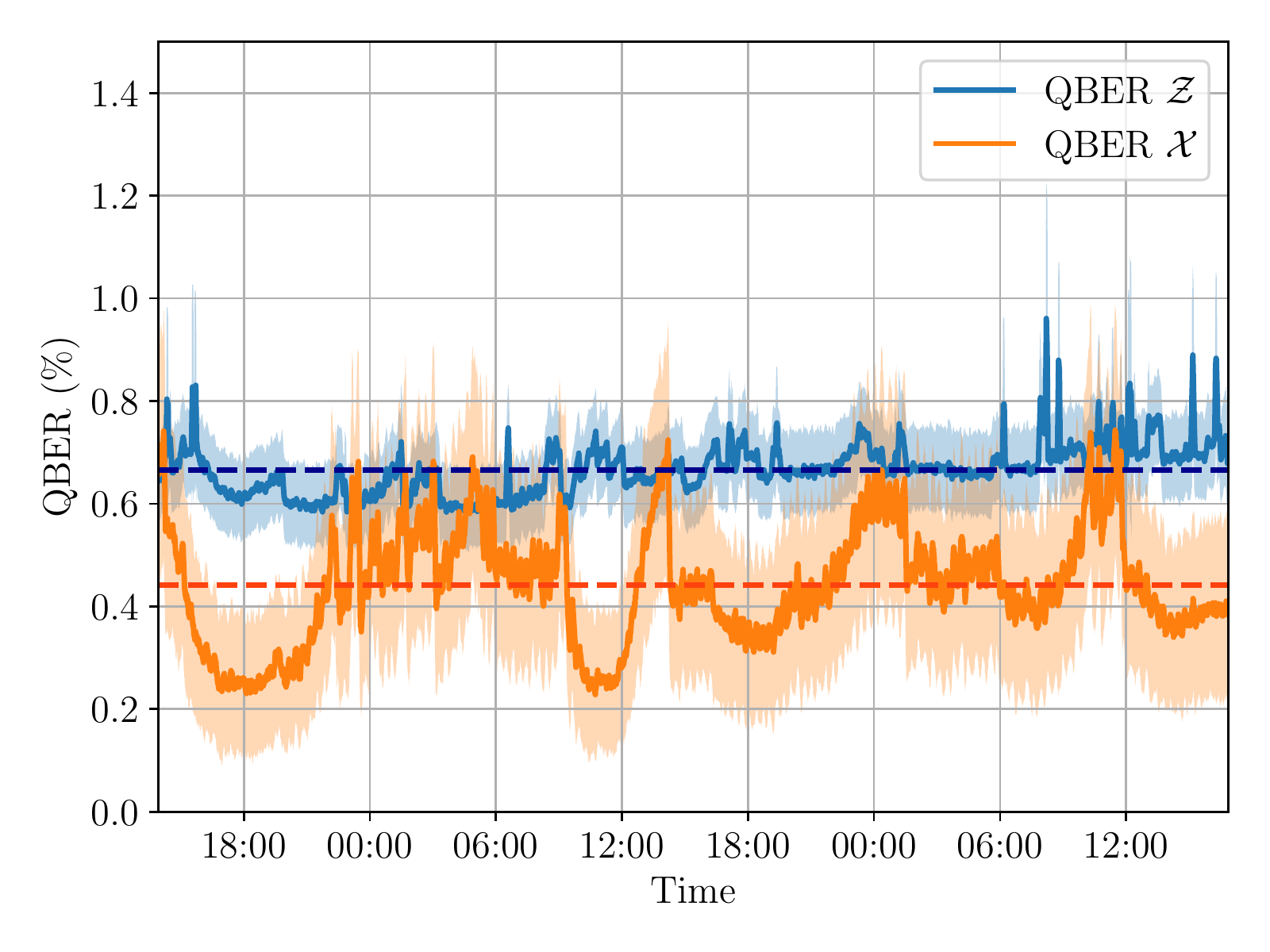} \\
    \includegraphics[width=.49\textwidth ]{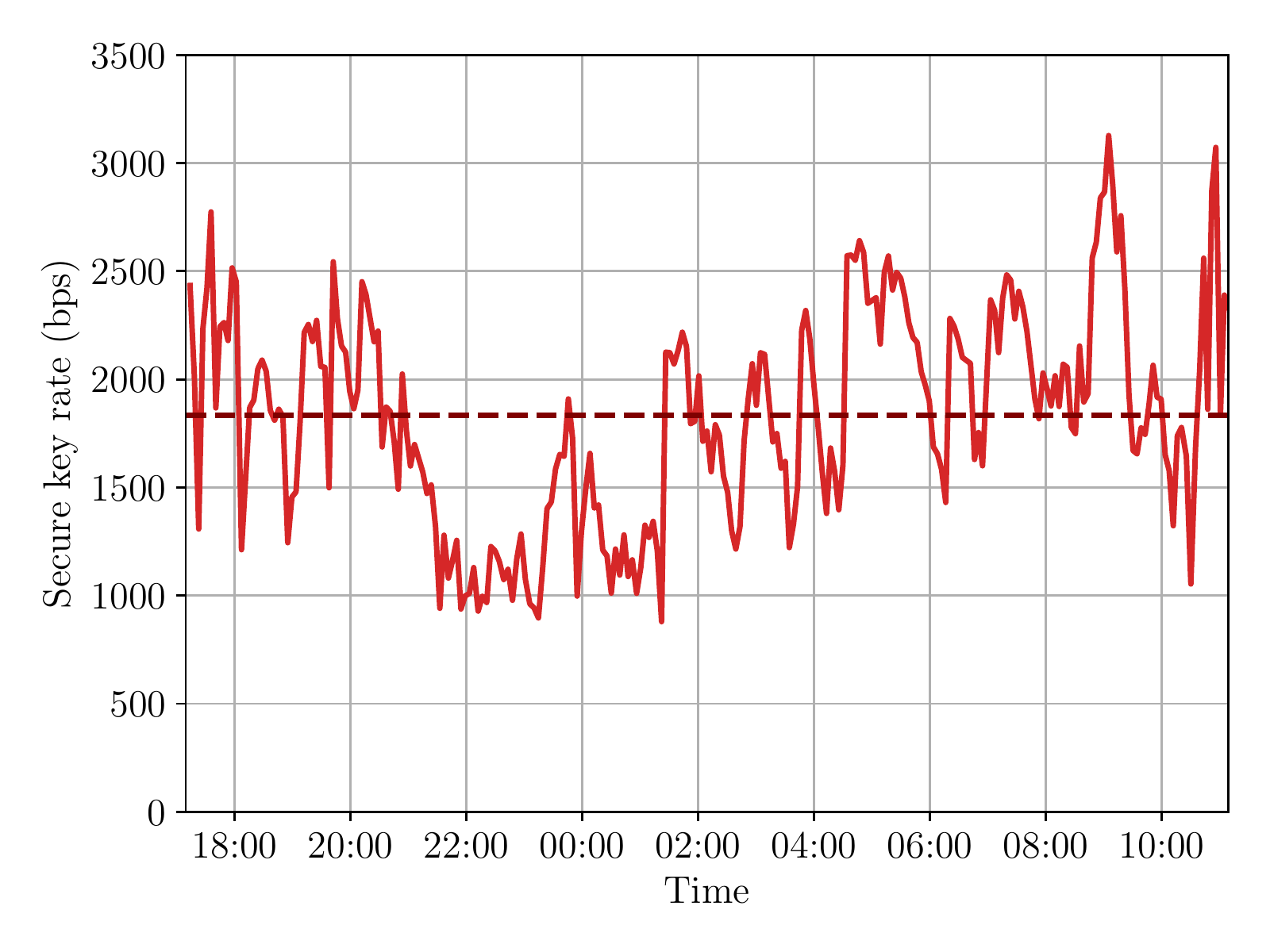}
    \includegraphics[width=.49\textwidth ]{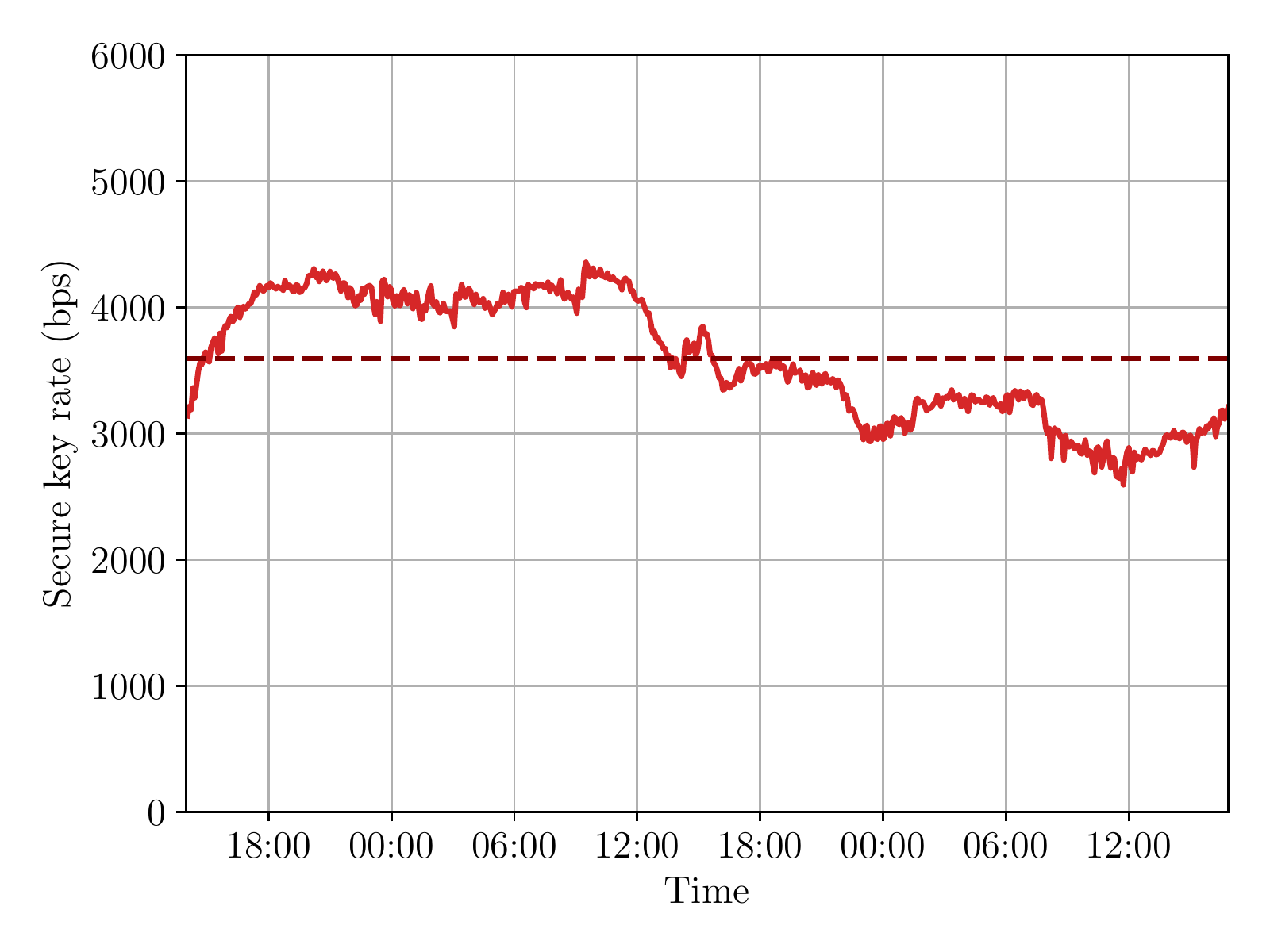}
    \caption{Summary of the performances of the two QKD systems in terms of QBER and SKR. The QBER is plotted with a rolling window of 5 minutes and the color bands represent $\pm 1$ standard deviations. Dashed lines are the average values. For Matera, they are $2.5\%$, $2.1\%$ and $1.9\text{ kbps}$ for the QBER in $\mathcal{Z}$ basis, QBER in $\mathcal{X}$ basis, and SKR respectively. For Oberpfaffenhofen, they are $0.7\%$, $0.4\%$ and $3.6\text{ kbps}$ for the QBER in $\mathcal{Z}$ basis, QBER in $\mathcal{X}$ basis, and SKR respectively.}
    \label{fig:summary_qkd}
\end{figure*}

\section{Last-mile QKD links}

\label{sec:last-mile}

The distribution of the shared key $k_{\rm sat}$ between the OGS and the PTF in MA, and between the OGS and the PTF in OP, was entrusted to two fiber-based QKD systems realizing the necessary “last-mile" connection to create the local keys $k_{\rm MA}$ and $k_{\rm OP}$. In this section, we explain in details how these two last-mile QKD links have been implemented at the two locations, and the performances --- in terms of quantum bit error rate (QBER), which is the mismatch of the signals sent by Alice and received by Bob, and secret key rate (SKR) --- achieved in our demonstration.

\subsection{Matera infrastructure}

The QKD system based in MA was similar to the one described in Ref.~\cite{Avesani:21}, which implements the $3$-state efficient BB84 protocol~\cite{Fung2006} with polarization modulation and 1-decoy technique~\cite{Rusca:18}. 
The transmitter is mainly composed of a laser at $1310$~nm, emitting pulses with a repetition rate of $50$~MHz, an intensity modulator, which allows setting the mean photon number needed for the decoys, and the polarization modulator based on the iPOGNAC scheme~\cite{Avesani2020}. It is worth noting that this is the first time the iPOGNAC scheme is used with a QKD signal in the O-band. Finally, the pulses are attenuated below the single-photon level with a variable optical attenuator before entering the fiber channel. The electronics is mainly composed by a System-on-a-Chip (SoC) with an FPGA and a CPU. Refer to Ref.~\cite{Stanco2022} for the full description of the SoC architecture.

After generation, the attenuated laser pulses enter the quantum channel traveling toward the receiver. The quantum channel is a $10$-km long standard single-mode telecom fiber: the total losses, including the extra losses at the transmitter and receiver interfaces, are $8.5$~dB.

The receiver decodes the states through a time-multiplexing scheme: although this scheme introduces an extra $3\text{ dB}$ of losses, it allows to improve compactness and reduces significantly the cost of the system. Indeed, the implemented system only requires one InGaAs/InP single-photon avalanche diode (SPAD), which in this case is a PDM-IR from Micro Photon Devices S.r.l., which provides 15\% quantum efficiency. 
The time tags corresponding to the photons arrivals are recorded by a quTAU, from qutools GmbH, time-to-digital converter and transmitted to a computer for data processing.

To further reduce architecture requirements, the QKD system exploits the Qubit4Sync algorithm~\cite{Calderaro2020} for the time synchronization between the transmitter and the receiver, avoiding the need for a dedicated system that distributes the clock reference.

In MA, the total key generation time during the demonstration lasted 18 hours and the system has generated $119$~Mb of secure key in that period. The performance of the QKD system used in MA are shown in the left panels of  Fig.~\ref{fig:summary_qkd}.

\subsection{Oberpfaffenhofen infrastructure}

\begin{figure}[!h]
    \centering
    \includegraphics[width=\columnwidth]{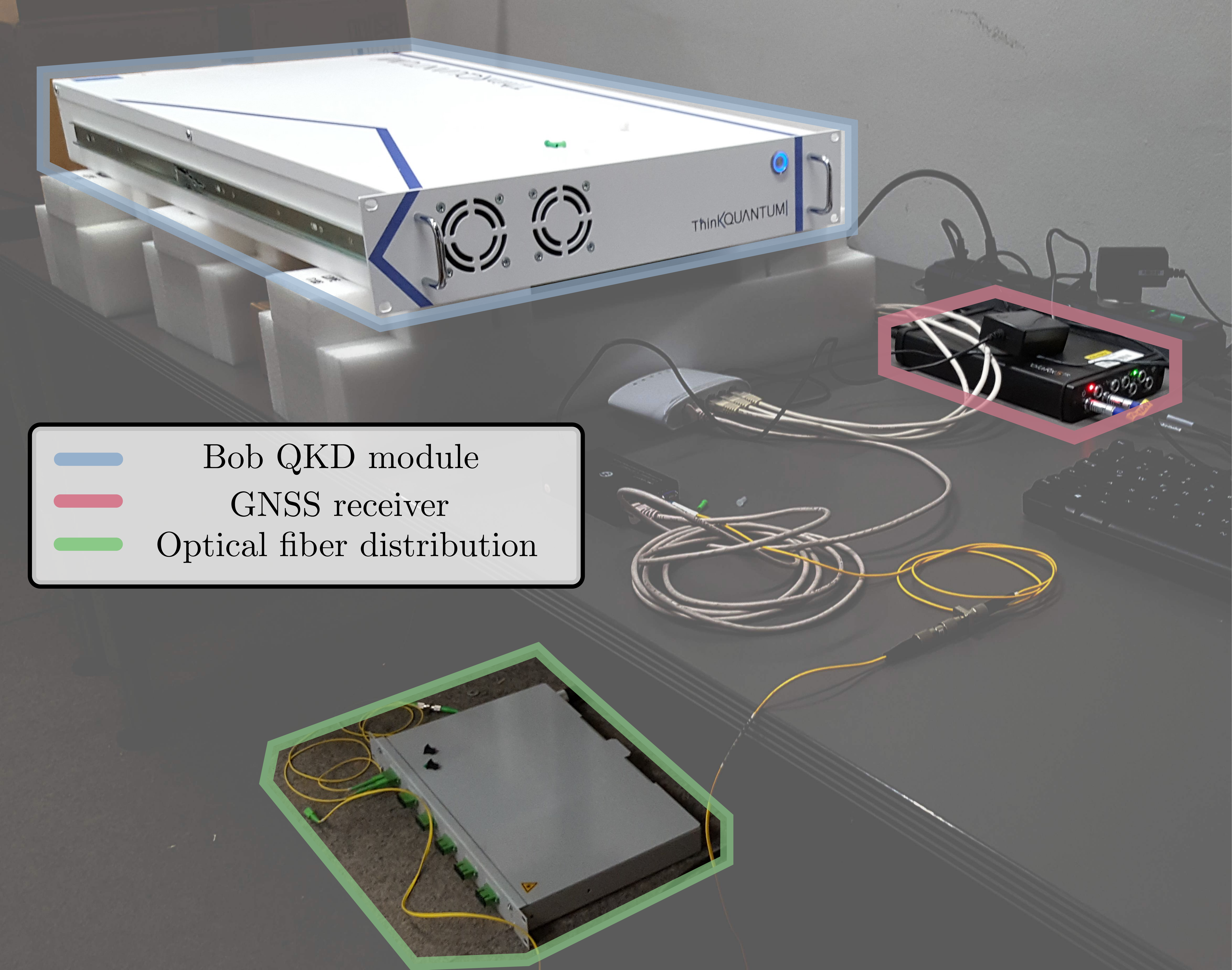}
    \caption{Photo of the experimental setup in OP. The Bob QKD module (ThinkQuantum) is based on the QUKY platform. The GNSS receiver is a commercial PolaRx5 receiver (Septentrio).}
    \label{fig:setup}
\end{figure}

For the specifications of the QKD system placed in OP, please refer to the QUKY platform developed by ThinkQuantum s.r.l. \cite{TQ}. As the system used in MA, the QUKY platform implements the 3-state 1-decoy efficient BB84 protocol via polarization encoding and the synchronization technique named Qubit4Sync, with the needed protocol stack, including authenticated transmission of support data and raw key post-processing (information reconciliation and privacy amplification) to yield a secure key

In the OP implementation, the quantum channel was a $500$-m long standard single-mode fiber to which extra losses were added, for a total amount of 12~dB.
The total working time was 51 hours and the system generated $660$~Mb of secure key. The performance of the QKD system used in OP are shown in the right panels of Figure~\ref{fig:summary_qkd}. A picture part of the setup OP is shown in Figure~\ref{fig:setup}.

\section{Satellite channel simulation}
\label{sec:satQKD}

To evaluate the pre-sharing of keys between the OGSs in MA and OP, which ideally originated from a satellite-to-ground QKD channel, we simulated a QKD protocol between a satellite transmitter and two receivers placed in our two ground stations. The purpose of the simulation is to estimate the average available secret key rate (SKR) that can be one-time-padded with the last-mile QKD key to propagate the pre-shared key to the end-nodes PTF-MA and PTF-OP. 

The two locations are already equipped with two telescopes with apertures $1.5$~m (MA) and $0.8$~m (OP). We assume that there are two QKD receivers for a $1550$~nm source that require a fiber-injection system to collect the signal from the telescopes up to the detectors.

The simulation goes as follows: first we propagate a satellite on a given orbit that allows it to pass over both the ground stations, then we compute for each pass the channel statistics to estimate the losses and the raw key accumulation on the ground stations, and finally we compute the SKR generation following standard finite-key security proofs~\cite{Rusca:18}.

\subsection{Orbit propagation}

\begin{figure}
    \centering
    \includegraphics[width=\columnwidth]{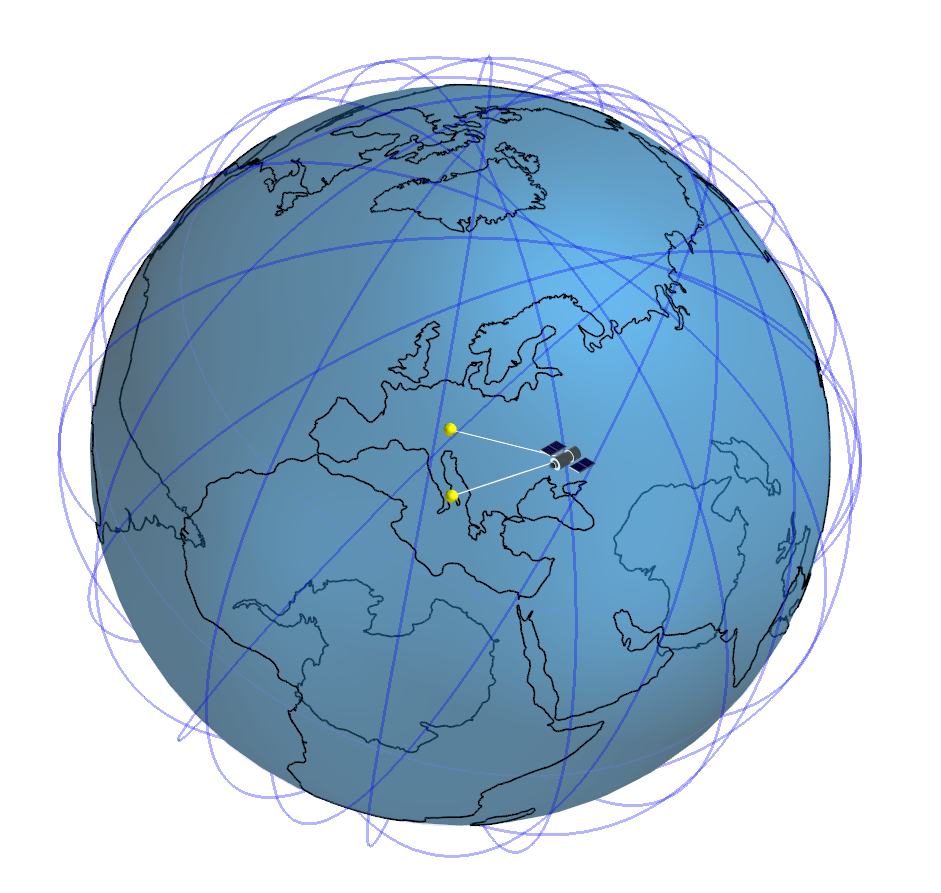}
    \caption{Propagation of the satellite orbit for one day of the experiment duration. The two dots represent the two ground station placed in MA, Italy, and OP, Germany.
    The blue line is the orbit through which the satellite has been propagated. The white segments indicate that an optical communication is available between the satellite and the ground stations.}
    \label{fig:orbit}
\end{figure}

We simulate a Low Earth Orbit (LEO) satellite at an altitude of $500$~km. The orbit has an inclination of $75.6$~degrees and a Right Ascension of the Ascending Node (RAAN) of $300.6$~degrees, passing over the two ground stations twice a day, and is equipped with a transmitter telescope having a diameter of $15$~cm, and a QKD system with a source at $1550$~nm. The satellite is propagated over a period of time, where the orbit is granuralized over steps of 1 second each. For each step, if a ground station is available to enable the optical quantum channel, a channel model analysis, described in the following sections, is made to compute the detection statistic. We consider only passs above 20 degrees of elevation, since at lower elevations the link quality is significantly hindered by atmospheric absorption and turbulence. In Figure~\ref{fig:orbit} an example of the simulated orbit is shown.

The average useful pass time per day for the two stations is $9.04$ minutes for MA and $10.20$ minutes for OP, with usually two passs per day. Note that, thanks to the wavelength choice, for which the atmosphere presents fewer losses and lower background noise, it is possible to generate a secure key even with the detections accumulated during daylight passs.

\subsection{Channel model}
For each point of the orbit we compute the channel efficiency from the satellite to the ground station, starting from the physical parameters of the transmitter, the receiver, and the channel. We consider only the average values at each point of the granularized orbit to calculate the channel efficiency.

\begin{figure}
    \centering
    \includegraphics[width=1.1\columnwidth]{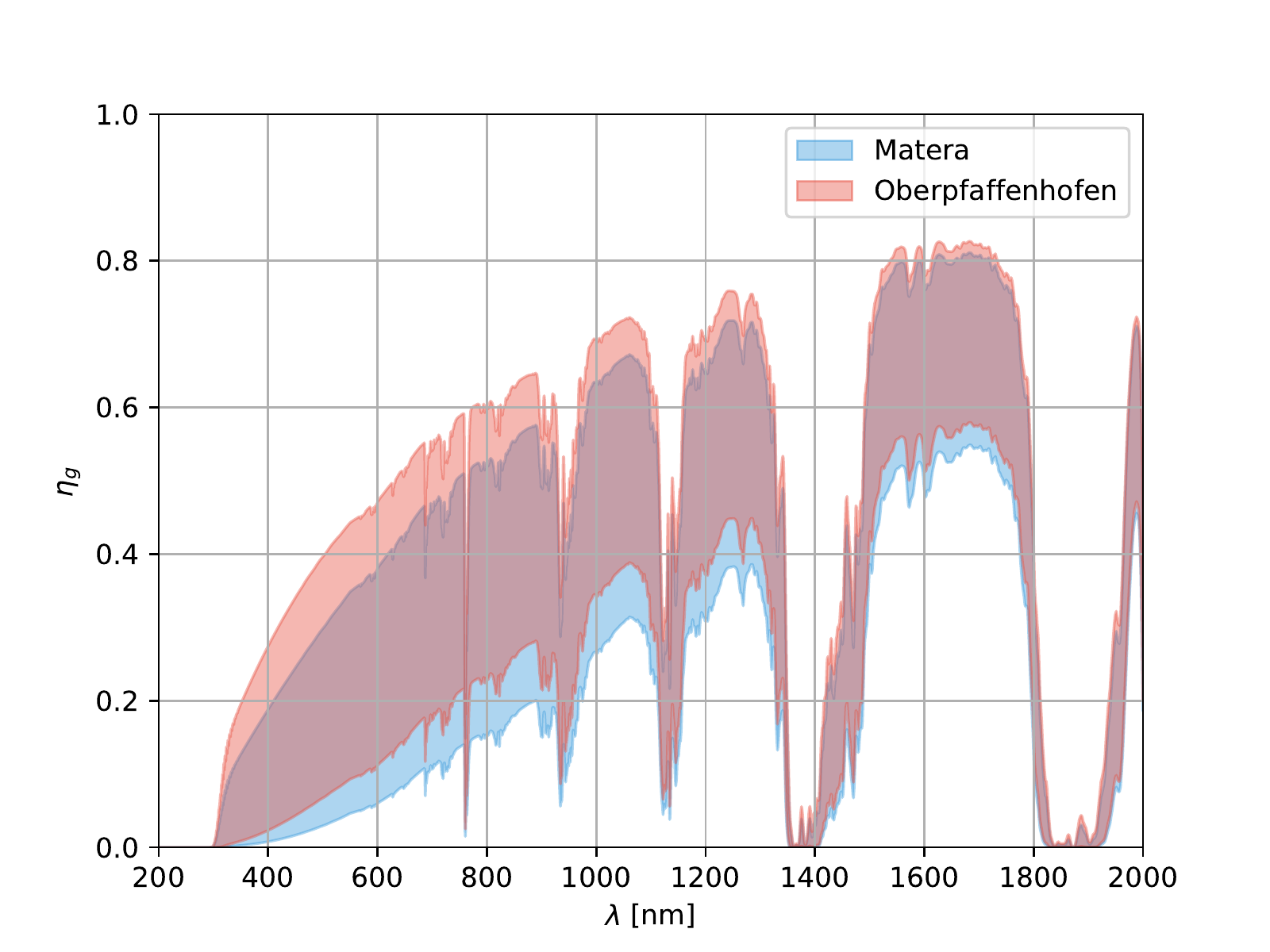}
    \caption{Atmospheric transmission from satellite to ground for the two sites in MA and OP, from zenith (higher transmission) to 20 degrees of elevation (lower transmission). The data was computed with LOWTRAN~\cite{lowtran} considering a mid-latitude winter atmosphere.}
    \label{fig:etaa}
\end{figure}

The channel efficiency $\eta$ is calculated as 
\begin{equation}
 \eta =   \eta_{a}  \eta_{g}  \eta_{f} \eta_{0}
\end{equation}
where $\eta_{a}$ is the atmospheric transmittance, caused by atmospheric scattering and absorption, retrieved from LOWTRAN~\cite{lowtran} (see Figure~\ref{fig:etaa}), $\eta_{g}$ is the geometric transmittance, caused by the finite size of the receiver telescope that clips the incoming beam, $\eta_{f}$ is the angular transmittance, caused by the finite angular acceptance of the receiving system and the pointing error, and $\eta_{0}$ takes into account all the fixed additional losses of the systems.

To compute $\eta_g$, we first propagate the optical beam from the transmitter aperture to the ground. The beam radius at the receiver $w_g$ depends on the diffraction-limited divergence of the Gaussian beam $\theta_{d}$, and on the divergence caused by the atmospheric turbulence $\theta_{t}$, which perturbs the beam wavefront. We assume that the angular pointing error of the transmitter is negligible.

Therefore, the total divergence $\theta$ is calculated as
\begin{equation}
    \theta = \sqrt{\theta_d^2 + \theta_t^2} = \sqrt{\left(\frac{\lambda}{\pi w_0}\right)^2 + \left(\frac{\lambda}{\pi \rho_0}\right)^2}
\end{equation}
where $\lambda$ is the QKD signal wavelength, $w_0$ is the Gaussian beam radius at the transmitter, and $\rho_0$ is the atmospheric coherence length of the spherical wave~\cite{Fante:75}. From the refractive-index structure constant $C_n^2$ one can compute $\rho_0$ for a satellite-to-ground path as 
\begin{equation}
    \rho_0 = \Bigg[1.46 k^2 R \int_0^1(1-\xi)^{\frac{5}{3}}C^2_n(\xi R) d\xi\Bigg]^{-\frac{3}{5}}
\end{equation}
where $k=2\pi/\lambda$ is the wavenumber of the photons, $R$ is the slant range to the ground station, the integral over $\xi$ considers the propagation of the beam through the path, and $C_n^2(\xi R)$ retrieves the $C_n^2$ associated to the height of the point $\xi R$. We employ the Hufnagel-Valley model for $C_n^2$, which has two free parameters: the value of $C_n^2$ at ground level and the average wind speed \cite{Andrews:05}.

The beam radius at the ground, exploiting paraxial approximation, is given by~\cite{Ricklin:02}
\begin{equation}
    w_g = \sqrt{w^2_0 + (\theta R)^2}
\end{equation}

Next, $\eta_{g}$ is calculated as the integral of a Gaussian beam over a circular aperture with an inner circular obscuration (as typical for Cassegrain configuration, where the secondary mirror partially occludes the primary mirror). This results in~\cite{Scriminich:22}
\begin{equation}
    \eta_{g} =  
    \exp\!{\bigg(\!-\frac{D^2_{\Rx,\textup{occ}}}{2w^2_g}\bigg)}
   -\exp\!{\bigg(\!-\frac{D^2_{\Rx}}{2w^2_g}\bigg)}
\end{equation} 
where $D_{\Rx}$ is the diameter of the primary mirror of the receiver telescope and $D_{\Rx, \textup{occ}}$ is the diameter of the (partially occluding) secondary mirror. Note that, in the limit $D_\Rx \ll w_g$, the geometric transmittance is approximately proportional to the telescope area, such that 
\begin{equation}
    \eta_{g} \approx (D^2_{\Rx} - D^2_{\Rx,\textup{occ}})/2w_{g}^2
\end{equation}
The angular transmittance $\eta_f$ is computed as the superposition of the field of view of the receiver telescope and the angular pointing error of the system: 
\begin{equation}
    \eta_f = 1 - \exp{\Big(-\frac{\theta_\Rx^2}{2 \alpha_\Rx^2}\Big)}
\end{equation}
where $\theta_\Rx$ is the intrinsic receiver half-field of view, and $\alpha_\Rx$ is the pointing error of the telescope.

In $\eta_0$ we take into account additional losses such as the losses due to the optics at the receiver, estimated to be around 3-5~dB, and the losses due to the fiber-injection of the signal, estimated to be around 8-10~dB for high-end adaptive optics systems \cite{Scriminich:22, Lim19}. Without the availability of measurements from different devices, we have decided to fix $\eta_0$ to $13$~dB.

\subsection{Secret key analysis}
We simulate the efficient BB84 protocol with one decoy state, following the security proof presented in Ref.~\cite{Rusca:18}. The security proof takes into account finite key effects, requiring to realize a previously chosen number of measurements before being able to get a secure key as an output of the protocol.

After the computation of the channel efficiency, and thus knowing the probability of receiving a photon from the satellite, we start the accumulation of photon counts on the receiver side, i.e., the quantum state measurement. For each point of the orbit, we can calculate the number of photons received at the detector during its associated time. The so-called \textit{raw key} is obtained by all the measurements in which Alice and Bob bases match, and after reaching the required key length for the finite-key analysis, we compute the secret key rate. We accumulate the raw key produced in different passs: this is a design choice which has the advantage of having a higher secure key generation rate, since for longer keys the finite key overheads are less prominent; on the other hand, it requires larger memory at the satellite's side and additional memory management for security reasons.
Due to our choice of using systems at a wavelength of $1550$~nm, and because we must couple the signal into a SMF to reduce the background noise, we can exploit high-end detectors to perform the measurement. The superconducting nanowire single-photon detectors (SNSPDs) can reach an efficiency $\eta_{det}$ higher than 90\% with an ultra-low noise of fewer than $100$~counts per second, and thus are the best choice for this type of application.
We have to consider the presence of errors during the protocol that will eventually lead to a mismatch between the secret keys shared by the two parties. The main sources of errors are random events that get registered by the detectors, coming from the background light of the channel and the detector itself (in the form of dark counts) and the incorrect encoding and decoding of the quantum state due to nonidealities of the quantum devices. While the other parameters are inputs of the simulation, the background light can be defined as the diffuse atmospheric radiance exploiting LOWTRAN~\cite{lowtran}. Other effects that impact the detection, for example the direct light coming from the sun's reflection on the satellite, are not considered.

In Table~\ref{tab:input} the most relevant parameters used in the simulations are shown.

\begin{table}[htb!]
    \centering
            {
    \begin{tabular}{ll}
    \toprule
    Quantity & Value \\
    \hline
    $\lambda$ & 1550~nm \\
    $\eta_{det}$ & 0.9 \\
    $w_0$ & 0.15~m \\
    $\eta_0$ & 13~dB \\
    $D^{\MA}_\Rx$ & 1.5~m \\
    $D^{\MA}_{\Rx,\textup{occ}}$ & 0.1~m \\
    $D^{\OP}_\Rx$ & 0.8~m \\
    $D^{\OP}_{\Rx,\textup{occ}}$ & 0.3~m \\
    $\theta_\Rx$ & \SI{6.25}{\micro\radian} \\
    $\alpha_\Rx$ & \SI{100}{\micro\radian} \\
    $C_n^2$ at ground level & $10^{-14}~\mathrm{m}^{-2/3}$ \\
    Wind speed & 21~m/s \\
    Satellite altitude & 500~km \\
    Satellite inclination & \ang{75.6} \\
    Satellite RAAN & \ang{300.6} \\
    Satellite argument perigee & \ang{84.38} \\
    Satellite mean anomaly & \ang{38.29} \\
    Source repetition rate & 500~MHz \\
    Coding error & 0.5~\% \\
    Dark count rate & 100~Hz \\
    Detector dead time & 10~ns \\
    Temporal jitter & 10~ps \\
    Finite key length & 100~Mb \\
    Secrecy parameter & $10^{-10}$ \\
    Correctness parameter & $10^{-15}$ \\
    \hline
    \end{tabular}
    }
    \caption{Relevant input parameters for the satellite QKD simulations.}
    \label{tab:input}
\end{table}

\subsection{Simulation results}
The total secret key generated over a full year in the two OGS is 5.42~Gb for MA and 1.71~Gb for OP, that is equal to an average SKR of 171.7~bps and 54.1~bps respectively. Since the encryption protocol implemented by the two encrpytion systems needs 256~bpm, the requirement is satisfied. The satellites generated an average of 1.55~Mpbs of raw key per pass, depending on the zenith angle.

If we consider the probabilities of having overcast days, that in the worst condition do not allow an optical communication, we are still above the threshold. From Ref.~\cite{forecast} we see that the overcast probability is 34.2~\% for MA and 55.3~\% for OP. The average SKRs become 113.01~bps and 24.17~bps instead.

\section{GNSS and clock data acquisition and data analysis}
\label{sec:GNSS}

The time offset between the clock in OP and the clock in MA is measured using the all-in-view method, as presented in Section~\ref{sec:common-view}~\cite{DeFraigne2003}. The propagation time $\tau_{\sat,\PTF}$ has to be estimated and then subtracted to perform a consistent comparison between the satellite clock and the PTF. The estimation $\wtau_{\sat,\PTF}$ of the true propagation time is obtained by adding together all the known effects that contribute to it, which are grouped into dynamic effects and static effects:
\begin{equation}
    \wtau_{\sat,\PTF} = \wtau_\textup{dynamic} + \wtau_\textup{static}
\end{equation}

The dynamically changing propagation time can be obtained as
\begin{equation}
    \wtau_\textup{dynamic} = \frac{\chi}{c} + \Delta t_\textup{rel} - \Delta t_\textup{tropo} - GD
\end{equation}
with
\begin{equation}
    \chi  = \left[P_\textup{IF} - \big\| \vec{x}_\textup{sat} - \vec{x}_\textup{rec,IF}\big\| - S \right]
\end{equation}
where $P_\textup{IF}$ is the ionosphere-free pseudorange between satellite and receiver, $\vec{x}_\textup{sat}$ is the position of the satellite in the International Terrestrial Reference Frame (ITRF) at the emission time, $\vec{x}_\textup{rec,IF}$ the ionosphere-free phase center of the receiver antenna in ITRF, $S$ is the Sagnac correction associated with Earth's rotation, $\Delta t_\textup{rel}$ are relativistic corrections, $\Delta t_\textup{tropo}$ the delay originated from the troposphere and $GD$ the group delay of the emitted signal by the satellite. More details can be found in Ref.~\cite{DeFraigne2003}. This quantity is usually computed for every satellite available during the measurement time and is displayed in the Common GPS GLONASS Time Transfer Standard (CGGTTS)~\cite{defraigne2015}. The analysis requires a 13-minute data collection without losing contact to the satellite. 

\begin{figure}
    \centering
    \includegraphics[width=.9\columnwidth]{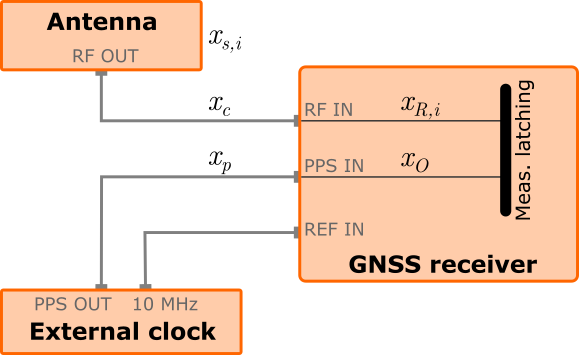}
    \caption{Schematic layout of the measurement delays of the system. 
    $x_{s,i}$: delay in the antenna for signal $i$; 
    $x_{R,i}$: delay in RF section of receiver for signal $i$; 
    $x_{C}$: delay in the RF cable (including amplifier and splitter); 
    $x_{p}$: delay in the PPS cable; 
    $x_{O}$: delay between PPS IN and internal receiver time reference. 
    See Ref.~\cite[Sec.~4.1]{PolaRx5} for further information.}
    \label{fig:septentrio}
\end{figure}

The static correction term is due to propagation delay in the PTF measurement system. These include the delays from the antenna, RF-cables, and within the GNSS-receiver:
\begin{equation}
    \wtau_\textup{static} = \frac{1}{c} (x_s + x_c + x_R - x_O - x_p)
\end{equation}
where the delay terms are explained in Figure~\ref{fig:septentrio}.

\subsection{GNSS data acquisition}

We have performed time-offset measurements between the OP PTF and the MA PTF. Both PTFs used a commercially available PolaRx5 receiver (Septentrio), connected to a choke ring multiband GNSS-antenna (Novatel GNSS-750, Leica AR25). The receiver is synchronized to an external $10$~MHz and $1$~PPS signals from the local PTF and can automatically compute the difference $\Delta  t_\textup{clock,sat}$ by applying standard ionospheric and tropospheric corrections. For simplicity, we did not calibrate the internal delays of the receiver system (antenna cables, RF-cables, etc.), although this shall be done when measuring the true offset between the two PTFs. The Septentrio receiver uses its own software for logging the GNSS-data (Septentrio Rx-control), which includes the navigation message for each observed satellite, the pseudoranges measured at both frequency bands, its power and signal-to-noise ratio and the Doppler shift~\cite{PolaRx5}.

The PolaRx5 receiver at each PTF continuously logged the GNSS-data and saved it in a lossless format called Septentrio Binary File (SBF). A new SBF file is created every day at 00:00, and the data are continuously appended to the current SBF file until the end of the day is reached. For time transfer and synchronization purposes, the SBF file is converted to CGGTTS format using the script \texttt{sbf2cggtts.exe} which is also provided by the Septentrio Rx-control software.

\begin{figure*}
    \centering
    \begin{minipage}{.49\textwidth}
        \textbf{MJD 59892, 9-Nov-2022} 
    \end{minipage}
    \begin{minipage}{.42\textwidth}
        \textbf{MJD 59893, 10-Nov-2022} 
    \end{minipage}
    \includegraphics[width=0.42\textwidth]{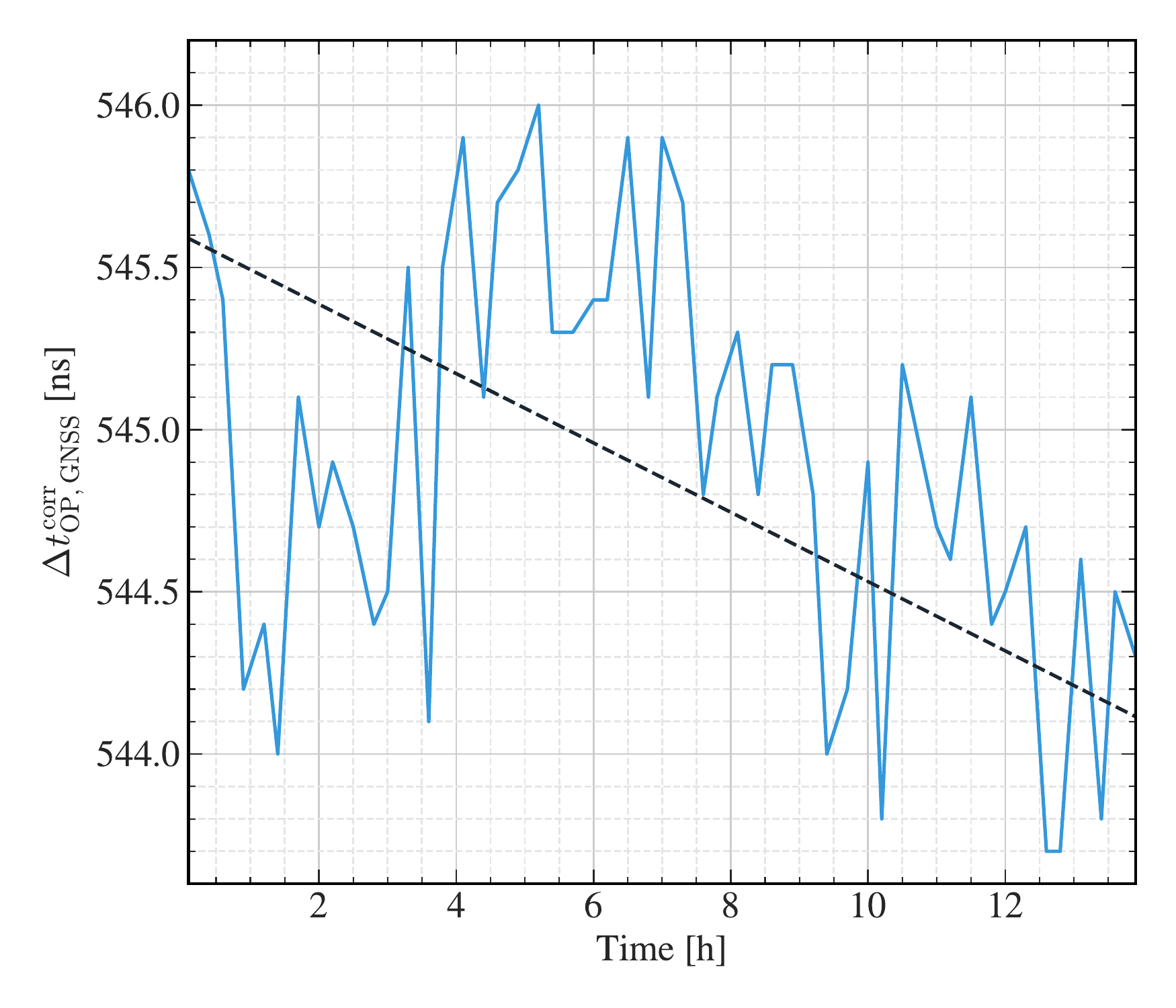} 
    \includegraphics[width=0.42\textwidth]{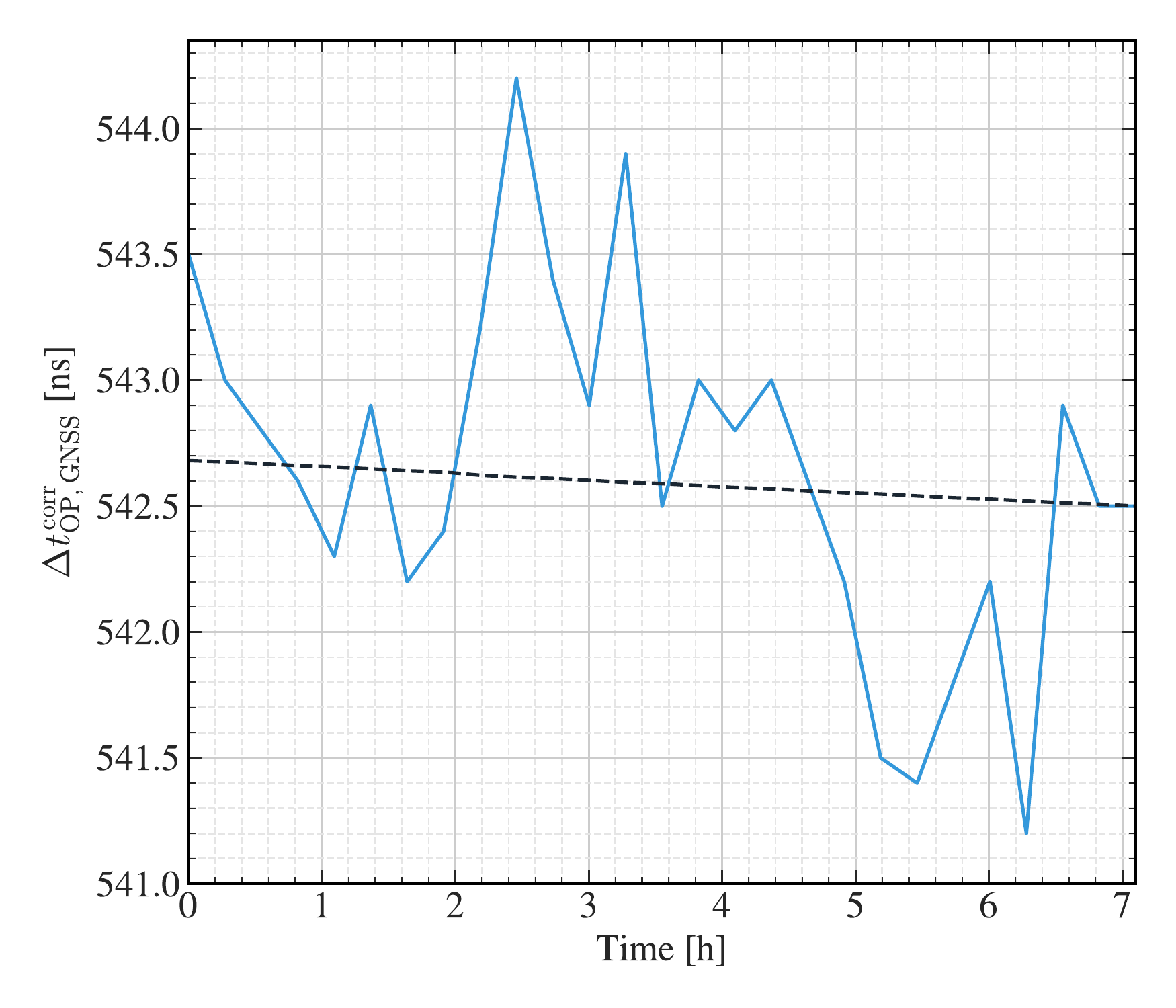} \\
    \includegraphics[width=0.42\textwidth]{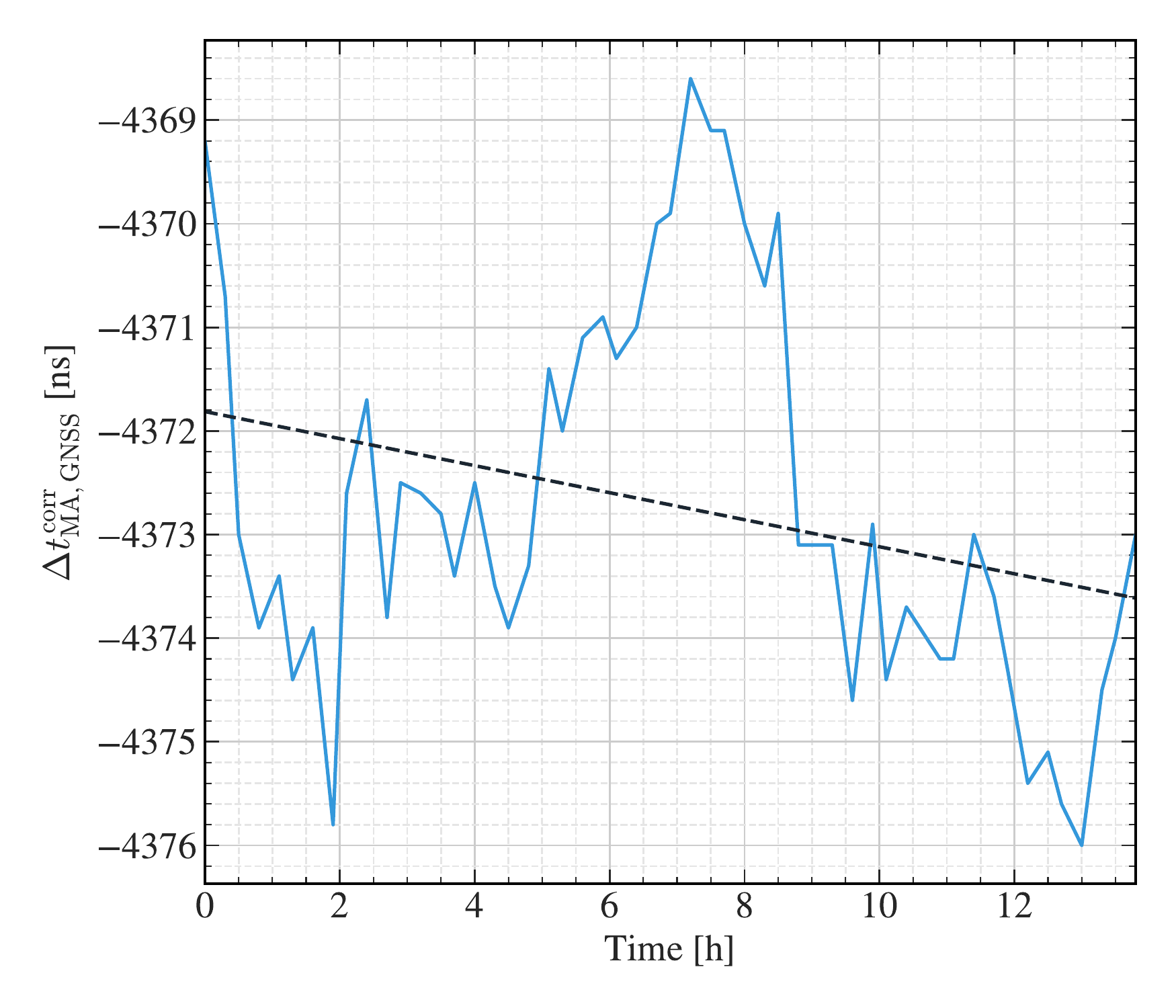}
    \includegraphics[width=0.42\textwidth]{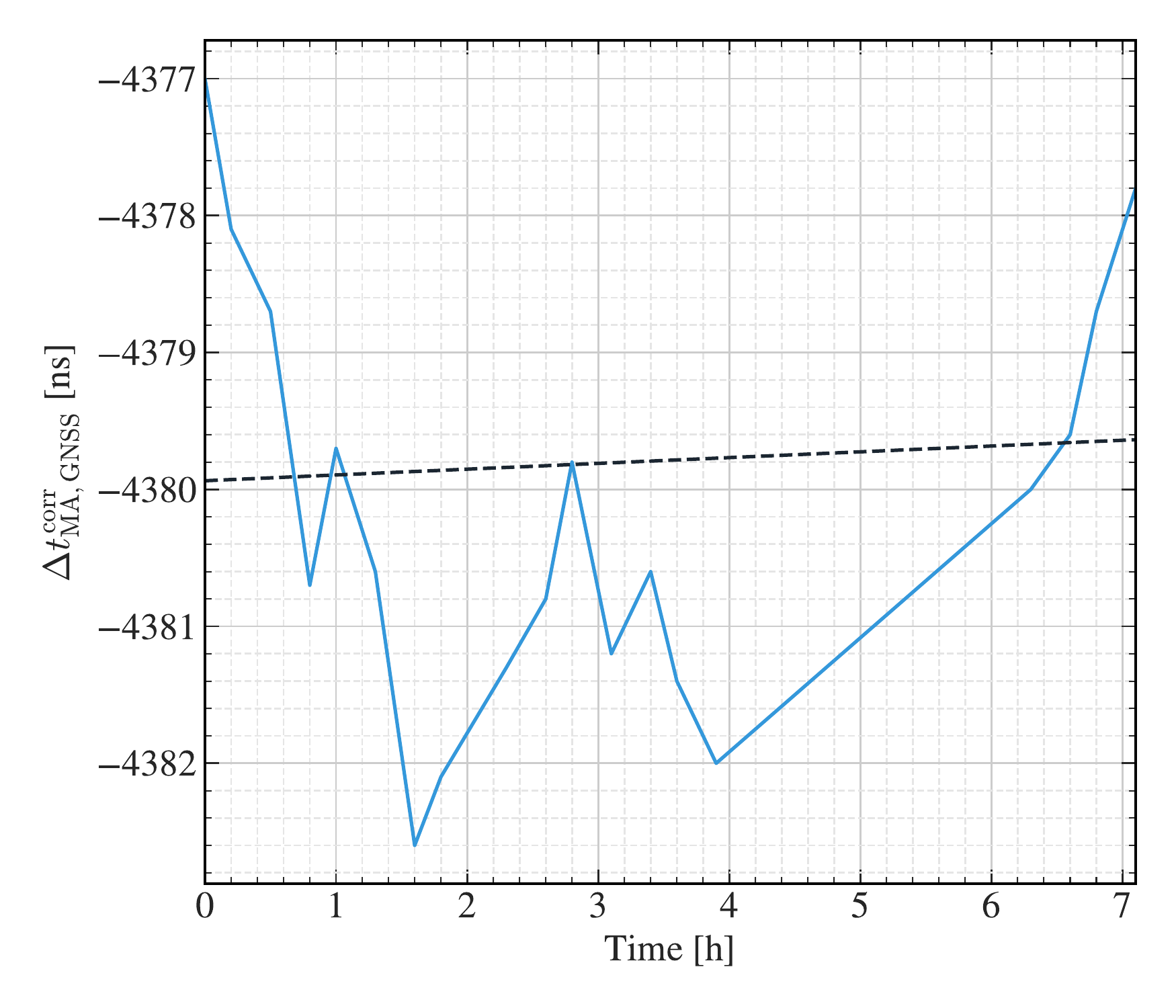} \\
    \includegraphics[width=0.42\textwidth]{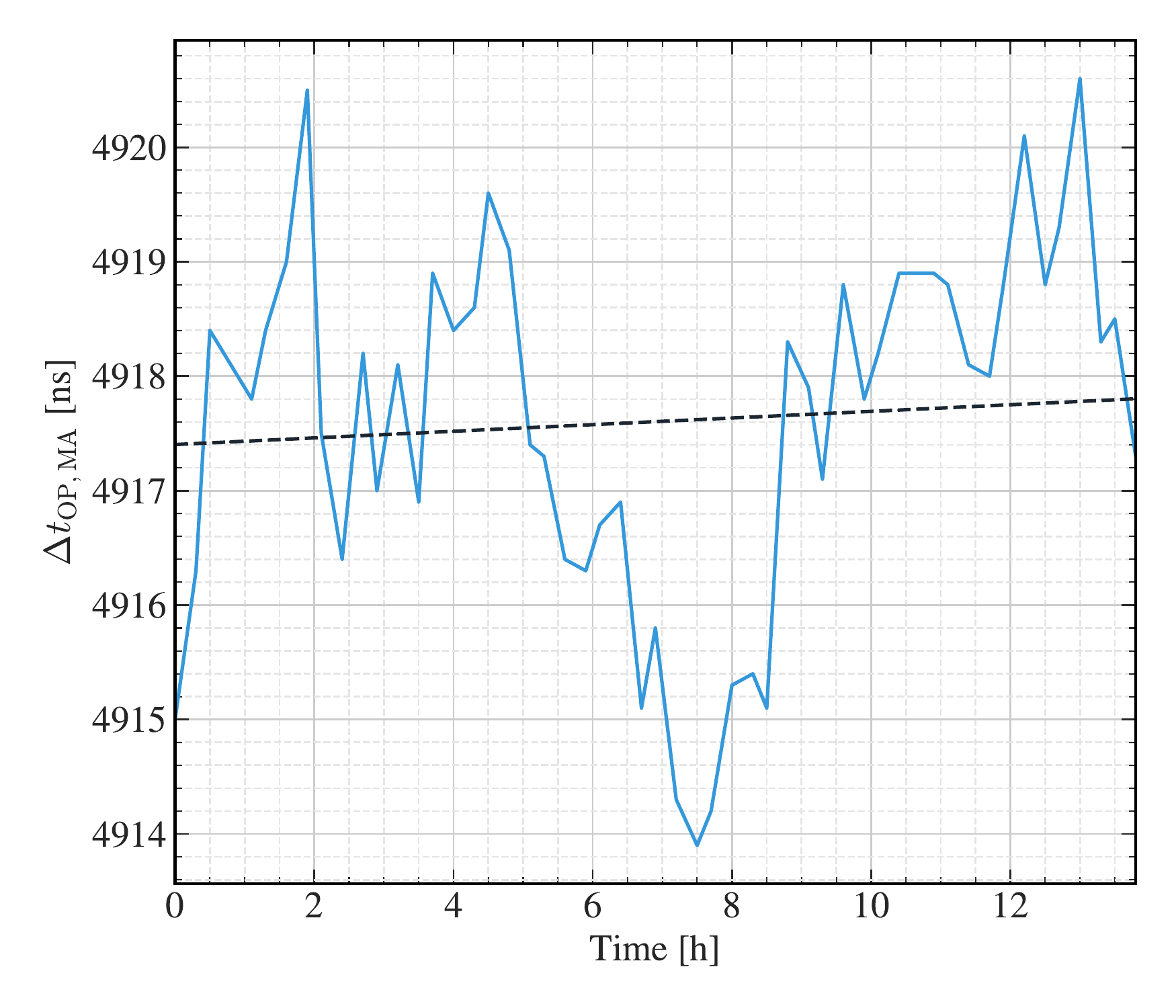}
    \includegraphics[width=0.42\textwidth]{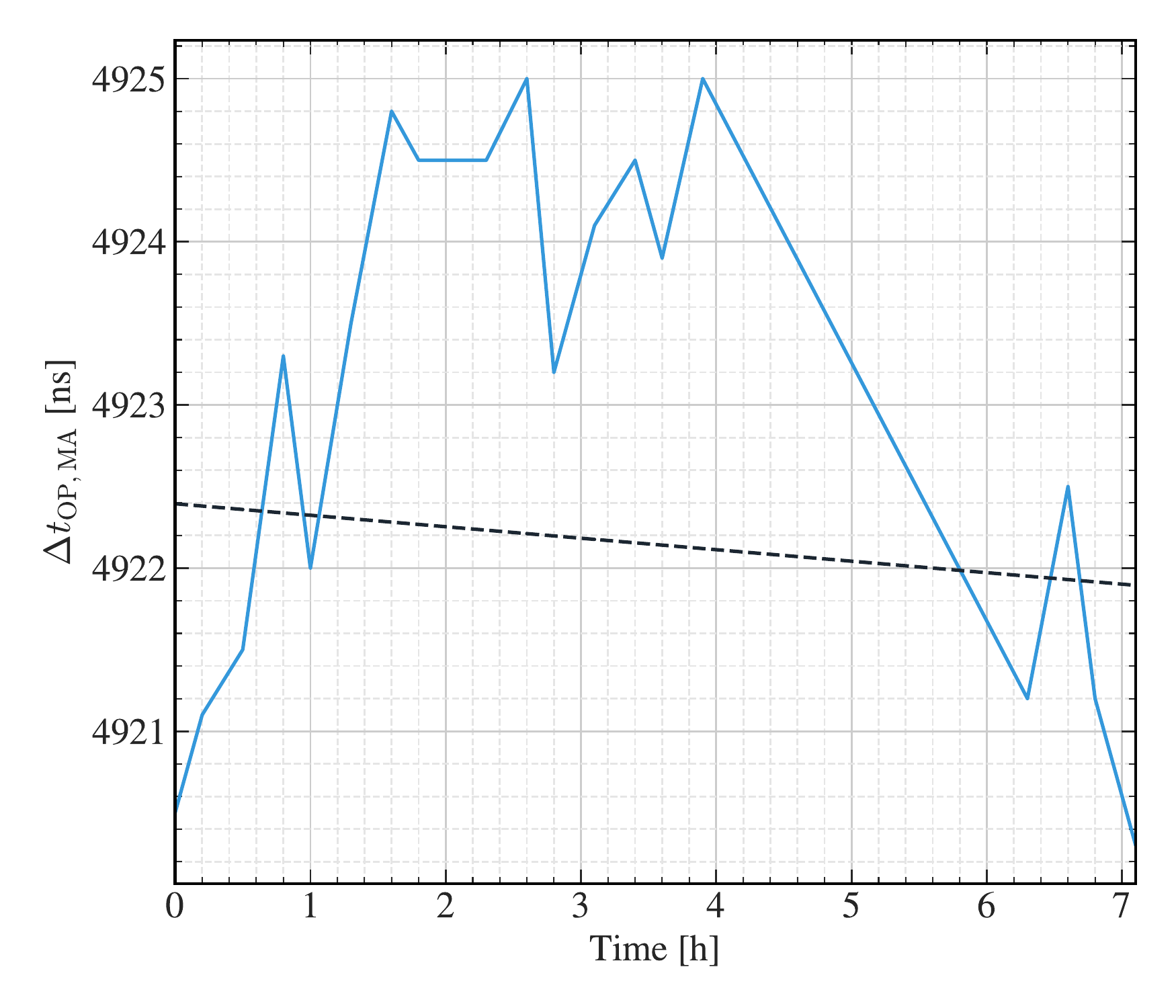} \\
    \caption{Time offsets measured during the experiment realization at MA and OP, and their difference $\Delta t_{\mathrm{OP,MA}}$. The first day of the measurement campaign (MJD 59892, 9-Nov-2022) is shown in the left column, with daily trend of the time offsets (dashed lines) $t_{\mathrm{OP}}=-2.6$~ns/day, $t_{\mathrm{MA}}=-3.1$~ns/day, and $\Delta t_{\mathrm{OP,MA}}=0.7$~ns/day. The second day of the measurement campaign (MJD 59893, 10-Nov-2022) is shown in the right column, with daily trend of the time offsets (dashed lines) $t_{\mathrm{OP}} = -0.6$~ns/day, $t_{\mathrm{MA}} = 1.0$~ns/day, and $\Delta t_{\mathrm{OP,MA}} = -1.7$~ns/day.}
    \label{fig:averages}
\end{figure*}

\subsection{Transfer of clock difference data via QKD-secured communication}
\label{sec:DataTransfer}

In our demonstration, the data is sent unidirectionally from MA to OP and stored locally, so that the time offset between the two clocks can be computed at the OP PTF; a bi-directional data transfer could be also implemented in future campaigns. Since the computation of the CGGTTS data requires at least 13 minutes of uninterrupted observation time, we decided to send a new update of the data once every 30 minutes.

The custom data transmission software opens the most recent SBF file and converts it to the CGGTTS format by calling the external \texttt{sbf2cggtts.exe} routine. A preliminary analysis of the data is performed, in which the $\Delta t_\textup{clock,sat}$ for every satellite during the current day observation is plotted. Furthermore, the median value of the measured clock offsets, as in Eq.~\eqref{eq:median}, is computed and plotted. 

The data collected from the MA lab are sent through the TCP/IP connection secured via QKD to the OP lab, see Section~\ref{sec:securing}. A tail-message is attached at the end of the file, displaying the time and date of the file sending, a checksum computed by adding the clock values of all satellites in view and the filename.

On the receiver site in OP, the received data from the MA lab are saved locally, and the same data analysis is performed as described above. A preliminary checksum analysis is computed to ensure that the data was transmitted consistently.

Next, a time window is determined to calculate the clock offsets. It is necessary that the CGGTTS data are available on both PTFs during the time window. This may not always be the case, for example, if one of the systems logging is stopped for a few hours or if it is necessary to reboot the system.

Finally, the difference $\Delta t_{\OP,\MA}^\textup{meas} \simeq t_{\OP} - t_{\MA}$ is calculated by subtracting the median values computed before, as in Eq.~\eqref{eq:deltaT2}.

\subsection{Difference in time offsets and clock synchronization in post-processing}

For this measurement campaign, we considered only the GNSS satellites from the European Galileo constellation, which are kept synchronized to the Galileo System Time (GST) by the ground station. By converting the SBF file into the CGGTTS format, it is possible to display the relative clock offset between the local PTF and a given Galileo satellite. Typically, there are at least 5 Galileo satellites that are observed at both laboratories at the same time. Some clock signals provide more accurate data for the all-in-view experiment than others depending on the specific satellite and its relative position to the PTF. In fact, different satellite clocks may have uneven performance, while local distortion effects such as multi-paths can become very large at low elevation angles above the horizon. Thus, a robust estimator of the clock offset between the PTF and the GST is obtained by taking the median value over all visible satellites at each observation time, as in Eq.~\eqref{eq:median}. 

As shown in the left column of Figure~\ref{fig:averages}, the relative drift between the OP PTF and the GST is about -2.6~ns/day, while the drift between MA PTF and GST is -3.1~ns/day (measured on MJD-59892 in which we have the largest number of measurements available). By computing the difference between the relative offset between the PTF clocks and the GST this is about 4917 ns on MJD-59892 with a relative drift of 0.7~ns/day which is, as expected, a very low value.

Once the clock offset data have been measured, a re-calibration procedure could be performed to synchronize the local and remote clock. In alternative, the measured clock difference can be saved locally and used to track the relative clock drift over time.

\section{Conclusions and outlook}
\label{sec:conclusion}

In this experiment, we have demonstrated the deployment of an integrated network allowing the transfer of time difference data secured by QKD. The experiment required the configuration and coordination of the operation of several subsystems, all of which were critical to the success of the experiment. These included atomic clocks both in MA and OP side, together with GNSS receivers, in order to perform an all-in-view time offset measurement; a fiber-based QKD system in MA and one in OP, each allowing the last-mile secure relaying of a quantum generated key; and finally the real-time secure transmission of the time difference data over an encrypted and authenticated internet connection. These subsystems were all simultaneously functional during the experimental campaign, leading to a successful demonstration of the use-case. 

We remark that the addition of a QKD security layer does not hinder the quality or timeliness of the time synchronization of the PTFs: the synchronization routines are typically performed only on a daily or weekly basis, since several hours of integration time are needed to measure a significant time offset between the local clocks~\cite{DeFraigne2003}. To see how this is possible, note that the time-critical part of the clock offset measurement is performed through GNSS signal acquisition, see Eq.~\eqref{eq:deltaT_loc2}. The clock offset between the two remote PTFs can then be determined by comparing the local time-difference data, as in Eq.~\eqref{eq:deltaT2}. This operation is not time-critical: data processing can be performed later to recover what was the clock offset at any given previous time.

Further developments of this use-case can be envisioned. Most prominently, it would be of uttermost importance to demonstrate the integration of real satellite QKD systems in the network for the long-haul relaying of the quantum generated keys. This may be done relatively soon, since several satellite experimental platforms are scheduled for launch in the upcoming years~\cite{sidhu2021}. Furthermore, it could be of interest to strengthen the security of the classical communication link, by replacing the AES-secured connection with information-theoretic secure encryption and authentication. On a longer-term perspective, it would be of high scientific (and practical) interest to be able to use QKD to secure all the steps in time synchronization, including the GNSS segment. In particular, using the methods demonstrated in Ref.~\cite{Dai2020} one could use QKD signals, together with precise and authentic satellite ranging data, to authenticate the validity of clock signals of future GNSS constellations.

\section*{Acknowledgements}
\label{sec:ack}

We would like to thank Michele Paradiso of e-Geos spa and Vincenzo Buompane of the Italian Space Agency for the collaboration and support at the Matera Laser Ranging Observatory.
This work was supported by: European Union’s Horizon 2020 research and innovation programme, project OpenQKD (grant agreement No 857156); Agenzia Spaziale Italiana, project Q-SecGroundSpace (Accordo n. 2018-
14-HH.0, CUP: E16J16001490001); European Union's PON Ricerca e Innovazione 2014-2020 FESR/FSC - Project ARS01-00734 QUANCOM.

\section*{Author contributions}
\label{sec:contr}

All authors contributed to the study conception and design.
Management and coordination responsibility for the research activity were performed by Davide Orsucci, Francesco Picciariello, Francesco Vedovato, Giuseppe Vallone, Paolo Villoresi, Amita Shrestha, and Florian Moll. 
The experiment realization and the data collection were performed by Francesco Picciariello, Francesco Vedovato, Davide Orsucci, Marco Avesani, Matteo Padovan, Luca Calderaro, Pablo Nahuel Dominguez, and Thomas Zechel.
The data analysis was performed by Francesco Picciariello, Francesco Vedovato, Matteo Padovan, Giulio Foletto, Pablo Nahuel Dominguez, and Thomas Zechel.
The software used was developed by Francesco Picciariello, Francesco Vedovato, Matteo Padovan, Giulio Foletto, Marco Avesani, Luca Calderaro, Pablo Nahuel Dominguez, and Thomas Zechel.
The following authors provided technical equipment for the realization of the experiment: Daniele Dequal, Ludwig Blümel, Tobias Schmidt, Luca Calderaro, Marco Avesani, Francesco Vedovato, Giuseppe Vallone, and Paolo Villoresi.
The first draft of the manuscript was written by Francesco Picciariello, Francesco Vedovato, Davide Orsucci, Matteo Padovan, Pablo Nahuel Dominguez, and Thomas Zechel.
Acquisition of the financial support for the project leading to this publication was performed by Florian Moll, Johann Furthner, Tobias Schmidt, Giuseppe Vallone, and Paolo Villoresi.
All authors read and approved the final manuscript.

\bibliographystyle{unsrt}

\end{document}